\documentclass[12pt]{article}
\pdfoutput=1
\usepackage{tikz}
\usetikzlibrary{matrix,arrows,decorations.pathmorphing,decorations.pathreplacing}
\usepackage{jheppub}
\usepackage{amsmath,amssymb,euscript,array,mathrsfs,appendix,ctable,marvosym}
\usepackage{mathtools}
\usepackage{arydshln}
\usepackage{todonotes}
\usepackage{graphicx}
\usepackage{titlesec}
\usepackage{color}
\usepackage{upgreek}

\def\vv{\upnu}

\def\Bk{{\boldsymbol k}}
\def\BH{{\cal H}}

\newcommand{\Tr}{\operatorname{Tr}}

\def\bra#1{\langle #1|}
\def\ket#1{| #1\rangle}

\newcommand{\EQ}[1]{\begin{equation}\begin{split} #1
\end{split}\end{equation}}

\title{Unravelling Cosmological Perturbations}
\author{Timothy J. Hollowood}
\affiliation{
Department of Physics, Swansea University,\\ Swansea, SA2 8PP, United Kingdom
}

\emailAdd{t.hollowood@swansea.ac.uk}

\abstract{We explain in detail the quantum-to-classical transition for the cosmological perturbations using only the standard rules of quantum mechanics: the Schr\"odinger equation and Born's rule applied to a subsystem.
We show that the conditioned, i.e.~intrinsic, pure state of the perturbations, is driven by the interactions with a generic environment, to become increasingly localized in field space as a mode exists the horizon during inflation. With a favourable coupling to the environment, the conditioned state of the perturbations becomes highly localized in field space due to the expansion of spacetime by a factor of roughly $\exp(-c \Delta N)$, where $\Delta N\sim 50$ and $c$ is a model dependent number of order 1. Effectively the state rapidly becomes specified completely by a point in phase space and an effective, classical, stochastic process emerges described by a classical Langevin equation. The statistics of the stochastic process is described by the solution of the master equation that describes the perturbations coupled to the environment.
}

\notoc
\begin{document}

\maketitle

\newpage

\section{Introduction}

It is breathtaking that quantum fluctuations \cite{F1,F2,F3,F4,F5,F6} in the inflating universe \cite{F7,F8,F9,F10,F11,F12} became the seeds of the structure in the universe and were imprinted as small fluctuations of the CMB. However, there seems to be a missing chapter in the standard story, namely, how the quantum fluctuations actually became classical. If one is only interested in probability distributions then the issue can be discussed only in terms of the degree of decoherence and so it sits there as the elephant in the room.

The goal here is to go beyond a description of a quantum process in terms of probabilities to one that can describe the trajectory of a single system. The classic ``Wigner's Friend" thought experiment illustrates the issues involved in a very simple setting, but one that is not meant to be at all realistic.\footnote{True macroscopic systems cannot be described by simple states like $\ket{F_\pm}$ because of entanglement with the environment that is being ignored here. In the simple toy model, it is the qubit that effectively decoheres the friend.} The friend measures a qubit initially in the state $c_+\ket{+}+c_-\ket{-}$ and according to the external observer, Wigner, the total state of the system is the entangled state $c_+\ket{+}\ket{F_+}+c_-\ket{-}\ket{F_-}$. More specifically, Wigner associates the reduced state $\rho=|c_+|^2\ket{F_+}\bra{F_+}+|c_-|^2\ket{F_-}\bra{F_-}$ to the friend. On the contrary, for the friend, Born's rule implies that their state is either $\ket{\psi}=\ket{F_\pm}$  with probability $|c_\pm|^2$, respectively.\footnote{The states $\ket{F_\pm}$ are the states in the generically unique decomposition of $\rho$ into an orthogonal ensemble or, equivalently, the eigenvectors of $\rho$.} So the state of the system depends on the frame of reference. The external observer, Wigner, describes the friend with the {\it unconditioned\/} state $\rho$ whereas in the friend's frame their state $\ket{\psi}$ is {\it conditioned\/}:
\EQ{
&\text{\bf W (unconditioned state):}~\qquad \rho=|c_+|^2\ket{F_+}\bra{F_+}+|c_-|^2\ket{F_-}\bra{F_-}\ ,\\[8pt]
&\text{\bf F (conditioned state):}\qquad\qquad \ket{\psi}=\begin{cases}\ket{F_+} & \text{prob}=|c_+|^2\ ,\\[8pt]\ket{F_-} & \text{prob}=|c_-|^2\ .\end{cases}
}
The two states are related via a stochastic average
\EQ{
\rho={\mathscr E}(\ket{\psi}\bra{\psi})\ .
\label{shh}
}
The key distinction between the two states is that the unconditioned state $\rho$ exhibits entanglement -- it is a mixed state -- while the conditioned state $\ket{\psi}$ is pure but random. So there is a {\it duality of perspective\/} between entanglement and randomness: $\rho\longleftrightarrow\ket{\psi}$, which gives rise to a form of {\it observer complementarity\/}.\footnote{By recognizing that the state depends on the frame of reference (or perspective, or context) one realizes a unification of {\it many worlds\/} and {\it Copenhagen\/} quantum mechanics.}

In the context of the cosmological perturbations, the unconditioned state $\rho$ is the one that is analysed in concrete models involving their interaction with some environment, consisting either of other fields or self interactions of the perturbations. After various approximations, the state $\rho$ satisfies a {\it master equation\/} that describes how the perturbations are decohered by the environment. This points to the fact that a classical description should be valid and probability densities can then be extracted from $\rho$. However, if one wants to describe how a single perturbation becomes classical, we need a description of the trajectory of the individual mode, in other words the state from the frame of the reference of the mode itself. This is the conditioned state constructed via the Born rule to satisfy \eqref{shh}. The formalism then decides whether the quantum-to-classical transition happens: does the state $\ket{\psi}$ becomes localized in phase space? The goal of this paper is to show that the conditioned state of the perturbations does become classical driven by interaction with the environment and the inflationary expansion.

It is well known that there are an infinite number of ways to write the solution of a master equation $\rho$ as a stochastic average as in \eqref{shh}, each known as an {\it unravelling\/}.\footnote{The terminology comes from the theory of {\it quantum trajectories\/} that describes the behaviour of a subsystem conditioned on the measurements made on it \cite{C1,JacobsSteck,W1,MW}.} However, there is a particular unravelling that follows from implementing the Born rule to a subsystem, the perturbations in the present context. This is the {\it Born unravelling\/} defined in \cite{Hollowood:2018ghk} and first described by Di\'osi \cite{Diosi1,Diosi2}.\footnote{See also \cite{Paz:1993tg,BH1,BH2,SH}. As shown in \cite{Paz:1993tg}, the Born unravelling also defines a set of {\it consistent histories\/} in the formalism of \cite{Griffiths,GellMann:1992kh,Omnes:1992ag}.} This has a phenomenology that is similar to  another unravelling, known as {\it quantum state diffusion\/} \cite{Gisin1,Gisin3,Gisin4} which has been widely studied as a description of the quantum-to-classical transition in \cite{Brun1,Brun2,Brun3,Rigo1,Gisin2,Bhattacharya:1999gx,BHJ1,GADBH,Habib:1998ai}. In both unravellings, the quantum-to-classical transition happens dynamically when the conditioned state becomes sufficiently localized that Ehrenfest's theorem applies and an effective description in terms of a position in phase space applies. In \cite{Hollowood:2018ghk} it was argued that the quantum-to-classical transition becomes a dynamical process that involves the following conceptual steps:
\begin{center}
{\scriptsize
\begin{tikzpicture}[scale=0.6]
\node at  (0,8) {system $+$ environment};
\draw[very thick] (0,8) ellipse (4cm and 0.7cm);
\node at (0,6) {master equation};
\draw[very thick] (0,6) ellipse (4cm and 0.7cm);
\node at (-5,4) {unconditioned state};
\draw[very thick] (-5,4) ellipse (4cm and 0.7cm);
\node at (-5,2.2) {probability densities};
\node at (-5,1.8) {Fokker-Planck equation};
\draw[very thick] (-5,2) ellipse (4cm and 0.7cm);
\node at (5,4.3) {conditioned state};
\node at (5,3.8) {quantum stochastic process};
\draw[very thick] (5,4) ellipse (4cm and 0.7cm);
\node at (5,2.3) {localization of state};
\node at (5,1.8) {point in phase space};
\draw[very thick] (5,2) ellipse (4cm and 0.7cm);
\node at (0,0.25) {classical stochastic process};
\node at (0,-0.25) {Langevin equation};
\draw[very thick] (0,0) ellipse (4cm and 0.7cm);
\draw[very thick,->] (0,7.3) -- (0,6.7);
\draw[very thick,->] (-5,3.3) -- (-5,2.7);
\draw[very thick,->] (5,3.3) -- (5,2.7);
\draw[very thick,->] (-2.2,5.4) -- (-2.6,4.6);
\draw[very thick,->] (2.2,5.4) -- (2.6,4.6);
\draw[very thick,<->] (-2.6,1.4) -- (-2.2,0.6);
\draw[very thick,->] (2.6,1.4) -- (2.2,0.6);
\node at (1.7,7) {Born-Markov};
\node at (5.8,5.1) {Born's rule: Born unravelling};
\node at (4.8,0.9) {Ehrenfest's Theorem};
\node at (-7.3,3) {semi-classical limit};
\node at (-10,0) {\phantom{.}};
\node at (-6,8) {$\boxed{\text{\small Quantum}}$};
\node at (-6,0) {$\boxed{\text{\small Classical}}$};
\end{tikzpicture}
}
\end{center}

\begin{itemize}
\item The unconditioned state $\rho$ of the subsystem of interest satisfies a master equation, within the Born-Markov approximation.
\item The conditioned state $\ket{\psi}$ (the state from the frame of reference of the subsystem) satisfies a particular unravelling of this master equation which takes the form of deterministic evolution with a non-linear, non-Hermitian, Hamiltonian, interspersed with stochastic jumps into orthogonal states (arising from applying the Born rule to each coherent interaction of the system with the environment).
\item Under favourable conditions, the dynamics of the conditioned state drive it to become localized on macroscopic scales and Ehrenfest's Theorem can be invoked. 
\item The localized state can be described by point in phase space (i.e.~a classical state) evolving according to the classical equations of motion plus stochastic noise, i.e.~a Langevin equation.
\item Finally, to bring things full circle, the Langevin equation has an associated Fokker-Planck equation whose solution is identified with the Wigner function of the unconditioned state in the semi-classical limit.
\end{itemize} 

The purpose of this work is to apply this formalism to the cosmological perturbations by considering their evolution according to the Born unravelling. We will argue that, with a suitable coupling to the environment, although the unconditioned state spreads out in field space when a mode exits the horizon during inflation, the conditioned state is driven to become increasingly localized in field space as a result of the expansion (just as described above). In the end the usual state analysed in the literature---the unconditioned state---becomes a probability density for the conditioned state that is effectively specified by a point in field space. We can follow the stochastic evolution of this state and find that it follows a random walk in field space once the mode under discussion has crossed the horizon. The CMB across the sky can be viewed as an ensemble of endpoints of the classical stochastic process.

The scalar curvature perturbations $\zeta $ are effectively described by a scalar field $\vv=\sqrt{2\varepsilon}a\zeta $, the Mukhanov-Sasaki variable, each Fourier mode of which is effectively a parametric oscillator whose Schr\"odinger equation looks like that of non-relativistic quantum mechanics:\footnote{We present the Schr\"odinger equation in a form that looks like a harmonic oscillator. The Hamiltonian is related to the Hamiltonian of the perturbations by a canonical transformation that just shifts the momentum $\uppi\to\uppi-(a'/a)\vv$. So before shifting, we have (classically) $\uppi=\vv'$ while after shifting $\uppi=\vv'-(a'/a)\vv=(\sqrt2\varepsilon a)\zeta'$.}
\EQ{
-\frac{\partial^2\psi}{\partial\vv^2}+\omega^2\vv^2\psi=2i\frac{\partial\psi}{\partial\tau}\ .
\label{yss}
}
Here, $\vv$ is identified with either of the real combinations $(\vv_\Bk+\vv_{-\Bk})/\sqrt2$ or $i(\vv_\Bk-\vv_{-\Bk})/\sqrt2$, of wave vector $\Bk$, and $\tau$ is the conformal time during inflation. The latter has negative values and approaches $\tau_\text{end}$ at the end of inflation. A mode exits the horizon when $k|\tau|\sim1$ and the modes of interest for the CMB and structure formation underwent $\Delta N\sim 50$ $e$-folds before the end of inflation, so $k|\tau_\text{end}|\sim e^{-\Delta N}$ for the modes of interest. Above, $a(\tau)$ is the scale factor. 

The power spectrum of the scalar curvature perturbations is simply related to the variance of the quantum mechanical problem,\footnote{In these formulae, we are ignoring the delta functions of the wave vector.}
\EQ{
\Delta_\zeta ^2=\frac{k^3}{2\pi^2}\langle\zeta\zeta\rangle=\frac{k^3}{4\pi^2\varepsilon a^2}\langle\vv^2\rangle\ .
}
In the above, 
\EQ{
\omega(\tau)^2=k^2-\frac{(a\sqrt{2\varepsilon})''}{a\sqrt{2\varepsilon}}\ ,
}
where $\varepsilon$ is a slow roll parameter. 
For present purposes, we will ignore slow roll effects and assume an exact de Sitter geometry during inflation for which $a=-1/(H\tau)$, for constant $H$, so $\omega^2=k^2-2/\tau^2$ and $\Delta_\zeta ^2=k^3H^2\tau^2\langle\vv^2\rangle/4\pi^2\varepsilon$.

The initial conditions of the mode are that at early times, $k|\tau|\gg1$, the mode sits in the ground state of the oscillator with $\omega=k$, the Bunch-Davies vacuum,
\EQ{
\psi(\vv,\tau)={\cal N}e^{-k \vv^2/2}\ .
}
Then as $\tau$ increases, at some point the mode crosses the horizon when $k|\tau|\sim1$. The oscillator becomes unstable and the state begins to spread out. We can describe the evolution in terms of the growth of the variance $V_\vv=\langle\vv^2\rangle$:\footnote{Note that for the unconditioned state $\langle\vv\rangle=0$.}
\EQ{
\frac14V'''_\vv+\omega\omega' V_\vv+\omega^2V'_\vv=0\ .
\label{uux}
}
In particular, as $\tau$ becomes small, the variance grows like
\EQ{
V_\vv\longrightarrow\frac1{2k^3\tau^2}\ ,
\label{rbb}
}
and the state is becoming highly squeezed in the conjugate direction. This implies that the perturbation $\zeta $ freezes since the power spectrum $\Delta_\zeta ^2=H^2/8\pi^2\varepsilon$ becomes independent of $k$ and $\tau$ at the end of inflation. 

The perceived problem is that the state, although spread out in the field direction, is still a pure state, so how can $|\psi(\vv,\tau)|^2$ be interpreted as a probability distribution in field space even though this is phenomenologically the right thing to do?

\section{Born unravelling}

In this section we digress to consider quantum mechanics. 
In quantum mechanics, an important r\^ole is played by observers, or more precisely {\it frames of reference\/} associated to subsystems. Consider the case where a subsystem $S$ is coupled to its environment ${\cal E}$ with a Hilbert space: $\BH=\BH_S\otimes\BH_{\cal E}$. As the subsystem $S$ interacts with ${\cal E}$, the total state will build up entanglement. An external observer, describes the composite system with the total state $\ket{\Psi}$, or if describing the subsystem $S$ specifically, then the reduced density matrix $\rho=\Tr_{\cal E}\ket{\Psi}\bra{\Psi}$. This is the {\it unconditioned\/} state. However, the state from the frame of the subsystem $S$ experiences entanglement as randomness according to Born's rule: so the state is one particular state of the ensemble determined by the eigen-decomposition of the reduced density matrix of $S$,
\EQ{
\rho=\sum_ip_i\ket{\psi_i}\bra{\psi_i}\ .
}
The eigenvalues $p_i$ are probabilities and the possible states $\ket{\psi_i}$ are orthonormal.\footnote{We can assume that in a generic situation there are no degeneracies. Any degeneracy that arises only occurs in an instance of time and does not lead to any non-analyticity in the following formalism.} From the point-of-view of the unconditioned state $\rho$, the $\ket{\psi_i}$ are distinct branches while the conditioned state only picks one of the states $\ket{\psi_i}$ determined randomly according to the probabilities. 

\begin{figure}[!h]
\begin{center}
\begin{tikzpicture}[scale=0.8]
\draw[very thick] (7,5) circle (1cm); 
\node at (7,5) {\large $S$};
\begin{scope}[scale=1.4]
\draw[very thick] (0,1) -- (10,1);
\draw[very thick] (0,2) -- (10,2);
\draw[very thick] (0.5,1) -- (0.5,2);
\draw[very thick] (1.5,1) -- (1.5,2);
\draw[very thick] (2.5,1) -- (2.5,2);
\draw[very thick] (3.5,1) -- (3.5,2);
\draw[very thick] (4.5,1) -- (4.5,2);
\draw[very thick] (5.5,1) -- (5.5,2);
\draw[very thick] (6.5,1) -- (6.5,2);
\draw[very thick] (7.5,1) -- (7.5,2);
\draw[very thick] (8.5,1) -- (8.5,2);
\draw[very thick] (9.5,1) -- (9.5,2);
\node at (1,1.45) {\small ${\cal E}_{a_{n-4}}$};
\node at (2,1.45) {\small ${\cal E}_{a_{n-3}}$};
\node at (3,1.45) {\small ${\cal E}_{a_{n-2}}$};
\node at (4,1.45) {\small ${\cal E}_{a_{n-1}}$};
\node at (5,1.45) {\small ${\cal E}_{a_{n}}$};
\node at (6,1.45) {\small ${\cal E}_{a_{n+1}}$};
\node at (7,1.45) {\small ${\cal E}_{a_{n+2}}$};
\node at (8,1.45) {\small ${\cal E}_{a_{n+3}}$};
\node at (9,1.45) {\small ${\cal E}_{a_{n+4}}$};
\end{scope}
\draw[very thick] (7,2.8) -- (7,4);
\draw[very thick,dotted] (1.3,2.8) -- (6.2,4.5);
\draw[very thick,dotted] (2.7,2.8) -- (6.25,4.4);
\draw[very thick,dotted] (4,2.8) -- (6.38,4.27);
\draw[very thick,dotted] (5.6,2.8) -- (6.6,4.1);
\draw[line width=1mm,->] (6,0.5) -- (4,0.5); 
\node at (7,0.5) {\small time};
\node at (8.5,3.6) {\small interaction};
\node at (2,4) {\small{entanglement}};
\end{tikzpicture}
\caption{\footnotesize  The ticker-tape paradigm for the environment that lies behind the Born-Markov approximation. In each time interval $\delta\tau$, the system $S$ interacts with a fresh bit of the environment and becomes entangled with it. These parts of the environment then disperse to leave only their entanglement and no further interaction. This continually decoheres the states of $S$ and gives rise to distinct branches. The conditioned state of $S$ follows one of these branches stochastically.}
\label{fig1} 
\end{center}
\end{figure}
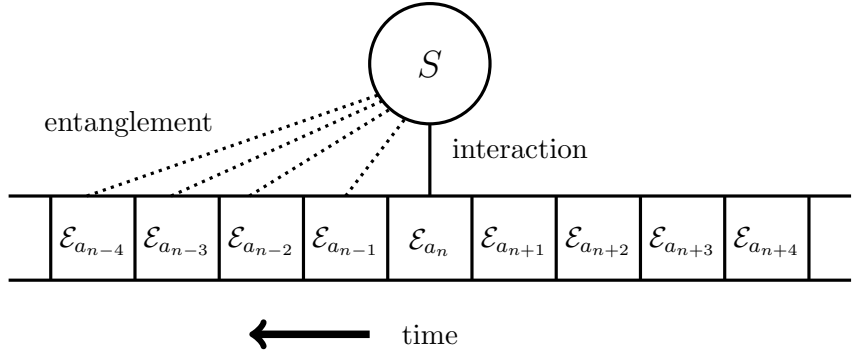

If $S$ interacts with ${\cal E}$ over a short time interval $\delta\tau$, then an initial pure state of $S$, $\ket{\psi}$ evolves, as a conditioned state, randomly to one of the $\ket{\psi_i}$ with probability $p_i$. One of the probabilities will be ${\cal O}(\delta\tau^0)$ while the others will be ${\cal O}(\delta\tau)$. In realistic situations, entanglement with the environment is dispersed rapidly and so we can model the environment as a series of subsystems ${\cal E}_a$. The subsystem $S$ interacts with a single component ${\cal E}_a$ for a short space of time $\delta\tau$ and interacts with the next component ${\cal E}_{a+1}$ in turn: see figure \ref{fig1}. After each interaction, entanglement is set up and the conditioned state evolves randomly as above. 
The Born-Markov approximation in this context means that the subsystem ${\cal E}_a$ interacts with $S$ over the short time interval and then never interacts with $S$ again so the branches are completely decoherent for all subsequent time. This approximation is known to be good for subsystems interacting with large environments which can rapidly disperse the entanglement with no back reaction.\footnote{Note that the assumption here makes the formalism more tractable but is not necessary.} Within this approximation, at longer time scales, we can effectively take $\delta\tau\to0$ and derive an autonomous differential equation---the master equation---for the unconditioned state of $S$, the density matrix $\rho$. This can always, on general grounds, can be written in Lindblad form as \cite{K1,GKS,L1}
\EQ{
i\frac{\partial\rho}{\partial\tau}=[H,\rho]+\frac i2\sum_i(2A_i\rho A_i^\dagger-A_i^\dagger A_i\rho-\rho A_i^\dagger A_i)\ ,
\label{mst}
}
where $H$ is an effective Hamiltonian and $A_i$ are the Lindblad operators.

Given the master equation \eqref{mst}, the dynamics of the conditioned state as described above is then determined as follows. The time interval $\delta\tau$ for the interaction with each element of the environment is infinitesimal at the level of the master equation. So during the time interval, there is a probability ${\cal O}(\delta\tau^0)$ that the state $\ket{\psi}$ evolves deterministically via an effective Schr\"odinger equation
\EQ{
\frac{\partial\ket{\psi}}{\partial\tau}=-iH_\text{eff}\ket{\psi}\ ,
\label{seq}
}
and a probability of ${\cal O}(\delta\tau)$ that it jumps into an orthogonal (but unnormalized) state $\ket{\psi_i}=J_i\ket{\psi}$, which defines the {\it branch creation operator\/} $J_i$. The normalization determines the rate of the jumps via
\EQ{
r_i=\bra{\psi}J_i^\dagger J_i\ket{\psi}\ ,
}
and the orthogonality condition requires
\EQ{
\bra{\psi}J_i\ket{\psi}=0\ ,\qquad \bra{\psi}J_i^\dagger J_j\ket{\psi}=r_i\delta_{ij}\ .
}

We now determine the branch creation operators explicitly.
The overall consistency condition is that
performing an average over the stochastic evolution of the conditioned state should give the unconditioned state, as in \eqref{shh}. Let us show this for the case when there is a single Hermitian branch creation operator $J$ (see \cite{Hollowood:2018ghk} for the general case). 
Over a short time interval $\delta\tau$, the unconditioned state should correspond to summing over the possible conditioned states, weighted with their probability, starting from $\rho=\ket{\psi}\bra{\psi}$.  With one branch creation operator $J$, there are two states to sum:
\EQ{
\begin{tikzpicture}
\node at (0,0) (a1) {$\rho=\ket{\psi}\bra{\psi}$};
\node at (7,1) (a2) {$\rho-iH_\text{eff}\rho\,\delta\tau+i\rho H^\dagger_\text{eff}\,\delta\tau$};
\node at (5.6,-1) (a3) {$J\rho J/r\ ,$};
\node[rotate=12] at (3,0.86) {\small $p=1-r\,\delta\tau$};
\node[rotate=-12] at (3,-0.86) {\small $p=r\,\delta\tau$};
\draw[very thick,->] (1.1,0.2) -- (4.8,1);
\draw[very thick,->] (1.1,-0.2) -- (4.8,-1);
\end{tikzpicture}
\label{ppp}
}
implies
\EQ{
\rho+\delta \rho=\big(1-r\delta\tau\big)\big(\rho-iH_\text{eff}\rho\,\delta\tau+i\rho H_\text{eff}^\dagger\,\delta\tau\big)+\frac1rJ\rho J\,(r\delta\tau)\ .
}
The $r=\bra{\psi}J^2\ket{\psi}$ in the denominator here is needed to normalize the state $J\rho J$.
Taking $\delta\tau\to0$, we can write this as 
\EQ{
\frac{\partial\rho}{\partial\tau}=-iH_\text{eff}\rho+i\rho H_\text{eff}^\dagger+J\rho J-r\rho\ .
\label{mst2}
}
Matching this to the master equation \eqref{mst} implies there is one Lindblad operator, and given the constraint that $J\ket{\psi}$ is orthogonal to $\ket{\psi}$, fixes the relation
\EQ{
J=A-\bra{\psi}A\ket{\psi}\ ,
\label{iss}
}
and the effective Hamiltonian that determines the evolution of the conditioned state in-between the jumps is
\EQ{ 
H_\text{eff}=H-i(J^2-r)/2\ .
\label{hef}
}
This is a non-Hermitian, non-linear, Hamiltonian and we will see that it has remarkable properties for the evolution of the conditioned state. Non-linearity arises because the operator $J$ depends implicitly on the state $\ket{\psi}$ via the expectation value in \eqref{iss}. In particular, the non-Hermitian term in \eqref{hef} has the tendency of driving the state towards a state that is annihilated by $J$: in other words, a state that is localized in the $A$ eigenbasis.

It is worth emphasizing that the dynamics of the conditioned state is defined for any quantum subsystem in \cite{Hollowood:2018ghk} in a completely general way the goes beyond the Born-Markov approximation. 

\section{Unravelling the perturbations}

There is a large literature discussing how cosmological perturbations became classical, including \cite{KP,KPS2000,KPS1998,Sakagami:1987mp,Lombardo:2005iz,Albrecht:1992kf,Polarski:1995jg,Lesgourgues:1996jc,Kiefer:1998qe,Kiefer:1998pb,Campo:2008ij,Brandenberger:1990bx,Kiefer:2006je,Martineau:2006ki,Burgess:2006jn,Kiefer:2008ku,Martin:2012pea,Burgess:2015ajz,Matacz:1996gk,Nelson:2017icq,Nelson:2017pmc}, but we should emphasize that in the present work we are considering a specific mechanism for how classical trajectories arise from quantum systems driven by decoherence by generic environmental interactions (Born unravelling). The question is whether this formalism can successfully explain how the perturbations appear as effectively classical, but stochastic, quantities. Note that in other approaches, it is simply assumed that the perturbation are classical stochastic variables, here we will show how that arises via the unravelling formalism.

It is unrealistic to suppose that the curvature perturbations evolve in isolation. We can expect that they are interacting with other fields and also coupled by self interactions. So the perturbations will form a subsystem of a much larger system. When the perturbations couple linearly to the environment then, subject to other approximations \cite{Boyanovsky:2015tba,Hollowood:2017bil}, 
the Schr\"odinger equation \eqref{yss} will be generalized to a master equation with a single Lindblad operator proportional to the field mode $\vv$: 
\EQ{
-\frac{\partial^2\rho}{\partial \vv^2}+\frac{\partial^2\rho}{\partial \vv^{\prime2}}+\omega^2(\vv^2-\vv^{\prime2})\rho
-i\sigma^2(\vv-\vv')^2\rho=2i\frac{\partial\rho}{\partial\tau}\ ,
\label{mse}
}
where $\rho=\rho(\vv,\vv')$ is the density matrix in the field basis and the term involving $\sigma=\sigma(\tau)$ describes the decoherence arising from coupling to the environment. 

On general grounds, for a generic model of this type, the decoherence term behaves as $\sigma^2\sim|\tau|^{-2}$ \cite{Burgess:2014eoa},\footnote{Written in terms of the perturbations $\zeta$, this is a term proportional to $a^3(\zeta-\zeta')^2\rho$ in the master equation written in terms of cosmic time $\partial\rho/\partial t$.} For example, in \cite{Boyanovsky:2015tba,Hollowood:2017bil}, the r\^ole of the environment is played by another scalar field $\Phi$, massless and conformally coupled, with an interaction $\vv\Phi^2$, other, dissipative, terms arise in the master equation; however, the form above has the dominant term that controls the decoherence as modes exit the horizon. Another concrete model of this form is the approach of \cite{Nelson:2016kjm} that identifies the environment with shorter wavelength modes of the curvature perturbation $\zeta$ mediated purely by gravitational self couplings. The dominant coupling is cubic with two derivatives of the form $\zeta(\partial\zeta)^2$ that couples a long wavelength mode, the system, to a pair of shorter wavelength modes, the environment. However, it is interesting, following \cite{Martin:2018zbe}, to analyse a more general class of models with the general scaling
\EQ{
\sigma^2=\alpha/|\tau|^{p-3}\ ,
\label{mbh}
} 
for constants $\alpha>0$. So the physical models have the special value $p=5$.  

The advantage of the simple model \eqref{mse}, is that the unconditioned state remains Gaussian, even in the presence of the decoherence term:
\EQ{
\rho={\cal N}\exp\big[-\Omega\vv^2-\Omega^*\vv^{\prime2}-\xi(\vv-\vv')^2\big]\ .
}
We can write the evolution in terms of the variance $V_\vv=1/(4\Omega_1)$, where $\Omega=\Omega_1+i\Omega_2$:
\EQ{
\frac14V'''_\vv+\omega\omega'V_\vv+\omega^2V_\vv'=\frac12\sigma^2\ ,
\label{uux2}
}
to compare with \eqref{uux}. As $\tau\to\tau_\text{end}$, we can express the decoherence correction to the power spectrum as a multiplicative factor $1+\Delta{\cal P}$, which behaves as $\Delta{\cal P}\sim \alpha k^{p-5}$, independent of $\tau$, if $p<8$, while there are potentially large corrections that behave as $\Delta{\cal P}\sim \alpha k^3|\tau|^{8-p}$, for $p>8$ \cite{Martin:2018zbe}.\footnote{The powers of $k$ here are just determined by power counting. A factor of $\alpha$ comes with $k^{p-5}$ and a factor of $|\tau|$ with $k$.}    The physical models with the special value $p=5$, leads to log corrections $\Delta{\cal P}\sim\log k$. 

The coupling to the environment leads to unconditioned state that becomes increasingly decoherent (mixed) as $\tau\to\tau_\text{end}$. A good way to visualize the decoherence of state is via the Wigner function
\EQ{
{\cal W}(\vv,\uppi)=\int_{-\infty}^\infty\frac{d\vv'}{2\pi}\,\rho(\vv+\vv'/2,\vv-\vv'/2)e^{i\vv'\uppi}={\cal N}\exp\big[-2\Omega_1\vv^2+\ell^2(\uppi+2\Omega_2\vv)^2\big]\ ,
\label{wff}
}
where $\ell=1/\sqrt{2\Omega_1+4\xi}$ is the coherence length that governs the fall off of the off-diagonal components of the density matrix:
\EQ{
\rho(\vv+\delta\vv,\vv-\delta\vv)\thicksim\exp\big[-\delta\vv^2/\ell^2\big]\ .
\label{rnn}
}
This is found to scale like
$\ell\longrightarrow|\tau|^{p/2-2}$ as $\tau\to\tau_\text{end}$. For the curvature perturbation $\zeta $, this would be a coherence length behaving as  $|\tau|^{p/2-1}$ which indicates that efficient decoherence by the end of inflation requires at least $p>2$.\footnote{For the curvature perturbations, and for the special value $p=5$, \eqref{rnn} becomes $\exp[-c\cdot a^3\delta\zeta^2]$. This matches the decoherence factor found in  \cite{Burgess:2014eoa} for a generic model with a linear coupling to the environment. It also matches the specific model of \cite{Nelson:2016kjm} based on gravitational self interactions.}
We can also quantify decoherence in terms of the area of the wave packet in phase space $\Delta$, where,
\EQ{
\Delta^2=4\langle\vv^2\rangle\langle\uppi^2\rangle-\langle\vv\uppi+\uppi\vv\rangle^2=1+2\xi/\Omega_1\ .
}
which is also the area of the Wigner ellipse
\EQ{
2\ell^{-2}\Omega_1\vv^2+(\uppi+2\Omega_2\vv)^2=(\pi\ell^2)^{-1}\ .
\label{vvt}
}
Initially the state is pure $\Delta=1$, but as it exits the horizon the area diverges like $\Delta\sim|\tau|^{2-p}$, for $p<8$, $|\tau|^{10-2p}$, for $p\geq8$. The area also determines the increase in the entanglement entropy of the mode 
\EQ{
S=\frac{\Delta+1}2\log\frac{\Delta+1}2-\frac{\Delta-1}2\log\frac{\Delta-1}2\ .
}
Hence, the rate of entropy production becomes fixed in cosmic time: $dS/dt\longrightarrow (p-2)H$, for $p<8$, and $(2p-10)H$, for $p\geq8$.

We can now unravel this master equation \eqref{mse} using Born unravelling. In the context of the perturbations, it follows from the master equation \eqref{mse} that the branch creation operator is $J=\sigma(\vv-\bar\vv)$. The deterministic part of the dynamics is the Schr\"odinger equation defined by the effective Hamiltonian \eqref{hef} 
\EQ{
-\frac{\partial^2\psi}{\partial\vv^2}+\omega^2\vv^2\psi-i\sigma^2(\vv-\bar \vv)^2\psi+ir\psi=2i\frac{\partial\psi}{\partial\tau}\ ,
\label{esh}
}
where $\bar\vv=\langle\vv\rangle$. On top of this, the state can jump randomly
\EQ{
\psi\longrightarrow \tilde\psi=\frac{\vv-\bar\vv}{\sqrt{V_\vv}}\psi\ ,
}
where $V_\vv=\langle (\vv-\bar\vv)^2\rangle$ is the field variance of the conditioned state, with a rate 
\EQ{
r=\langle J^\dagger J\rangle=\sigma^2 V_\vv\ .
\label{gcc}
}
Note that $\tilde\psi$ is normalized and orthogonal to $\psi$.
The Schr\"odinger equation is non-linear because $\bar\vv$ depends upon the state $\psi$ and this is why it can lead to localization of the state. Intuitively the non-Hermitian term proportional to $\sigma^2$ is trying to drive the state to be an eigenstate of $\vv-\bar\vv$: in other words, localized in field space.

\pgfdeclareimage[interpolate=true,width=10cm]{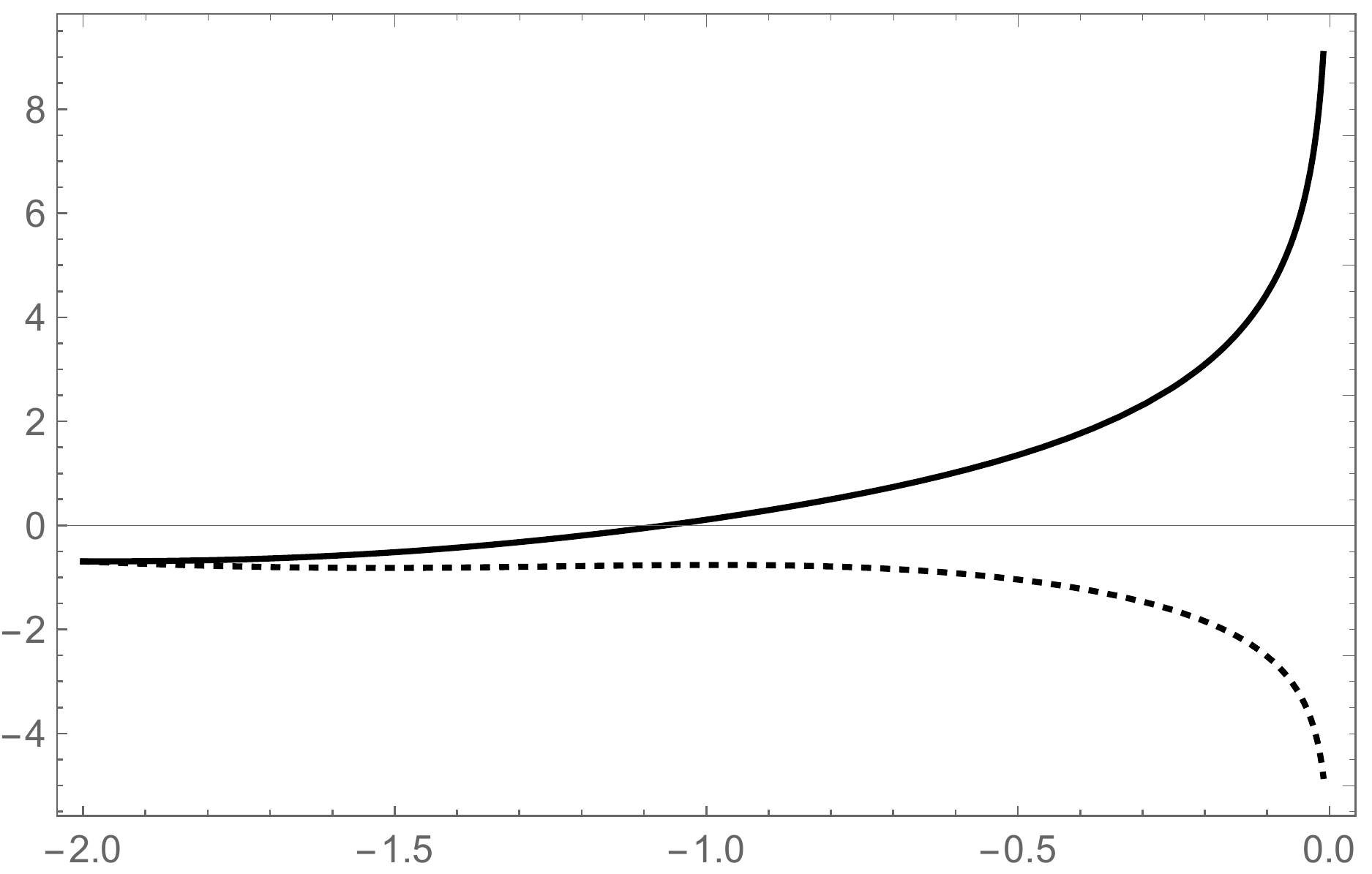}{p1}
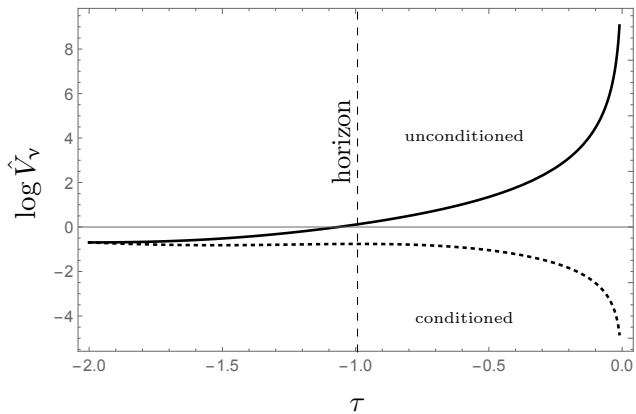
\begin{figure}[!h]
\begin{center}
\begin{tikzpicture}[scale=0.78]
\pgftext[at=\pgfpoint{0cm}{0cm},left,base]{\pgfuseimage{p1}} 
\node at (5.18,-0.5) {\small$\tau$};
\node[rotate=90] at (-0.5,3.3) {\small$\log\hat V_\vv$};
\draw[dashed] (5.18,0.4) -- (5.18,6.25);
\node[rotate=90] at (4.9,4) {\footnotesize horizon}; 
\node (a1) at (7,4.1) {\tiny unconditioned}; 
\node (a2) at (7,1) {\tiny conditioned}; 
\end{tikzpicture}
\caption{\footnotesize  The evolution of the variance $\hat V_\vv$ of the both the unconditioned and conditioned state for the deterministic evolution with $k=1$ as it exits the horizon at $k|\tau|\sim1$ with $p=5$. The localization of the conditioned state is clearly evident.}
\label{fig2} 
\end{center}
\end{figure}

We can solve the Schr\"odinger equation \eqref{esh} with a Gaussian ansatz:
\EQ{
\psi={\cal N}\exp\big[-\Xi(\vv-\bar\vv)^2+iP(\vv-\bar\vv)\big]\ ,
\label{gcs}
}
for ${\cal N}$, $\Xi$, $P$ and $\bar\vv$, all functions of $\tau$.\footnote{In the following we set $k=1$ but note that the momentum dependence can easily be re-introduced by noticing that from \eqref{esh}
$\psi(\vv)=f(k^{1/2}\vv,k\tau,\sigma/k)$.} By substituting the ansatz \eqref{gcs} into \eqref{esh}, one can readily derive the following equations for the field variance $\hat V_\vv$, the momentum $\uppi=-i\partial/\partial\vv$ variance $\hat V_\uppi=\langle(\uppi-\bar\uppi)^2\rangle$ and 
covariance $\hat C_{\vv\uppi}=\frac12\langle\{\vv-\bar\vv,\uppi-\bar\uppi\}\rangle$:
\EQ{
\frac{d\hat V_\vv}{d\tau}&=2\hat C_{\vv\uppi}-4\sigma^2 \hat V_\vv^2\ ,\\[5pt]
\frac{d\hat V_\uppi}{d\tau}&=-2\omega^2\hat C_{\vv\uppi}+\sigma^2-4\sigma^2\hat C_{\vv\uppi}^2\ ,\\[5pt]
\frac{d\hat C_{\vv\uppi}}{d\tau}&=\hat V_\uppi-\omega^2\hat V_\vv-4\sigma^2\hat V_\vv\hat C_{\vv\uppi}\ .
}
The hat on a quantity, indicate that it refers to an expectation with the Gaussian state evolving according to the effective Schr\"odinger equation \eqref{esh}. We also have the relations
\EQ{
\hat V_\uppi=\frac{1+(4\sigma^2 \hat V_\vv^2+\hat V'_\vv)^2}{4\hat V_\vv}\ ,\qquad
\hat C_{\vv\uppi}=2\sigma^2\hat V_\vv^2+\frac12\hat V'_\vv\ ,
}
and the area of the conditioned state $\Delta=1$ because the state is pure.

A relevant question is, how do $\hat V_\vv$, $\hat V_\uppi$ and $\hat C_{\vv\uppi}$ behave towards the end of inflation $\tau\to\tau_\text{end}$. With the model coupling \eqref{mbh}, we find the asymptotic scaling
\EQ{
p>5:\qquad\hat V_\vv\longrightarrow\frac1{2\sqrt\alpha}|\tau|^{(p-3)/2}\ ,\qquad\hat V_\uppi\longrightarrow\sqrt\alpha |\tau|^{(3-p)/2}\ ,\qquad \hat C_{\vv\uppi}\longrightarrow\frac12\ ,
\label{dcc}
}
In the regime, $3\leq p<5$, the scaling is more complicated and we simply write the exponents:
\EQ{
3\leq p\leq5:&\qquad\hat V_\vv\longrightarrow c_1 |\tau|^{p-4}\ ,\qquad\hat V_\uppi\longrightarrow c_2 |\tau|^{p-6}\ , \qquad\hat C_{\vv\uppi}\longrightarrow c_3 |\tau|^{p-5}\ .
\label{dcc2}
}
It is interesting that the dividing line between the two branches has $p=5$, precisely the special value. 

The behaviour \eqref{dcc} and \eqref{dcc2} means that the Gaussian wave packet becomes very narrow in field space, compared with the variance of the unconditioned state $V_\vv\sim|\tau|^{-2}$ \eqref{rbb}, as the mode exits the horizon and the end of inflation is approached: see figure \ref{fig2}. This is exactly what is needed to ensure that the conditioned state is effectively classical and is the main result of this paper. 

\pgfdeclareimage[interpolate=true,width=10cm]{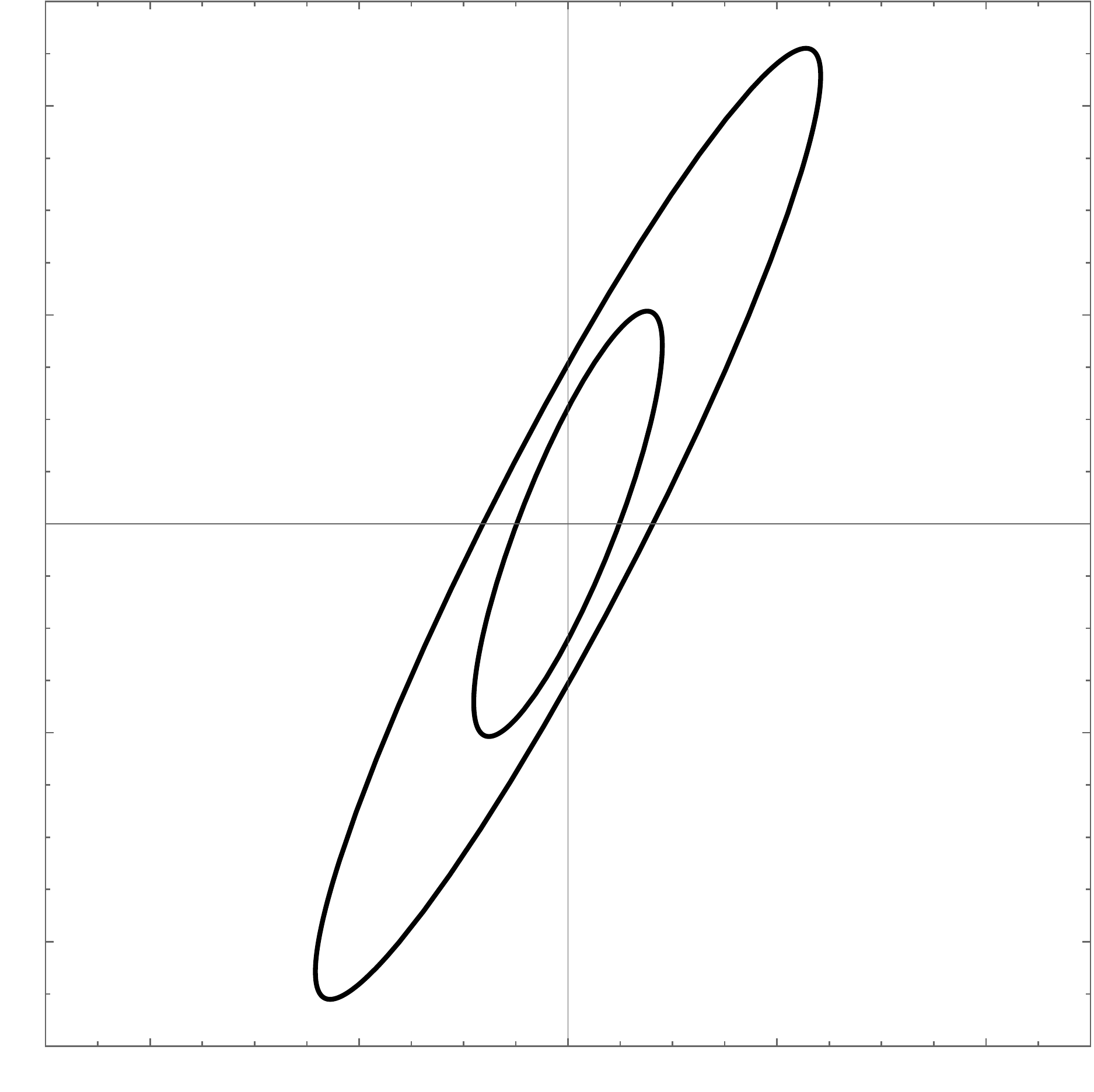}{p2}
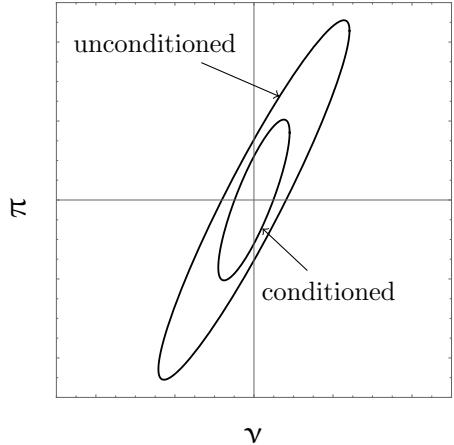
\begin{figure}[!h]
\begin{center}
\begin{tikzpicture}[scale=0.55]
\pgftext[at=\pgfpoint{0cm}{0cm},left,base]{\pgfuseimage{p2}} 
\node at (5.18,-0.5) {$\vv$};
\node[rotate=90] at (-0.5,5) {$\uppi$};
\node (a1) at (2.8,9) {\footnotesize unconditioned}; 
\node (a2) at (7,3) {\footnotesize conditioned}; 
\draw[->] (a1) -- (5.8,7.7);
\draw[->] (a2) -- (5.4,4.5);
\end{tikzpicture}
\caption{\footnotesize  The Wigner ellipses of the unconditioned and conditioned states (the latter centred at the origin) soon after the mode exits the horizon and squeezing begins.}
\label{fig3} 
\end{center}
\end{figure}

An important question is how localized does the conditioned state become relative to the unconditioned state? This can be answered by comparing the sizes of the associated Wigner ellipses: see figure \ref{fig3}. The one for the unconditioned state is written in \eqref{vvt} while for the Gaussian conditioned state \eqref{gcs}, we have
\EQ{
4\Xi_1^2\vv^2+(\uppi+2\Xi_2\vv)^2=2\Xi_1/\pi\ .
}
One can verify that there is an important distinction between the regimes $p\leq 5$ and $p>5$. In both cases, the ratio of the semi-major axis of the conditioned to unconditioned state scales to 0 as $\tau\to\tau_\text{end}$. However, it is only for $p>5$ that the ratio of the semi-minor axis of the conditioned to unconditioned state scales to 0 as $\tau\to\tau_\text{end}$. For $p\leq5$, the ratio remains fixed.
So it is only for $p>5$, that the unconditioned state effectively looks point-like and it becomes consistent to construct a coarse grained description that just involves specifying the position of the conditioned state in phase space. We can also see this distinction in the intercept of the Wigner ellipse along the momentum axis. For the unconditioned state, this equals the inverse coherence length $1/\ell$ and scales like $|\tau|^{2-p/2}$. The same intercept for the unconditioned state is $1/\sqrt{2\hat V_v}$ which scales in the same way, for $p\leq 5$, but which scales like $|\tau|^{(3-p)/4}$, for $p>5$. 

\pgfdeclareimage[interpolate=true,width=10cm]{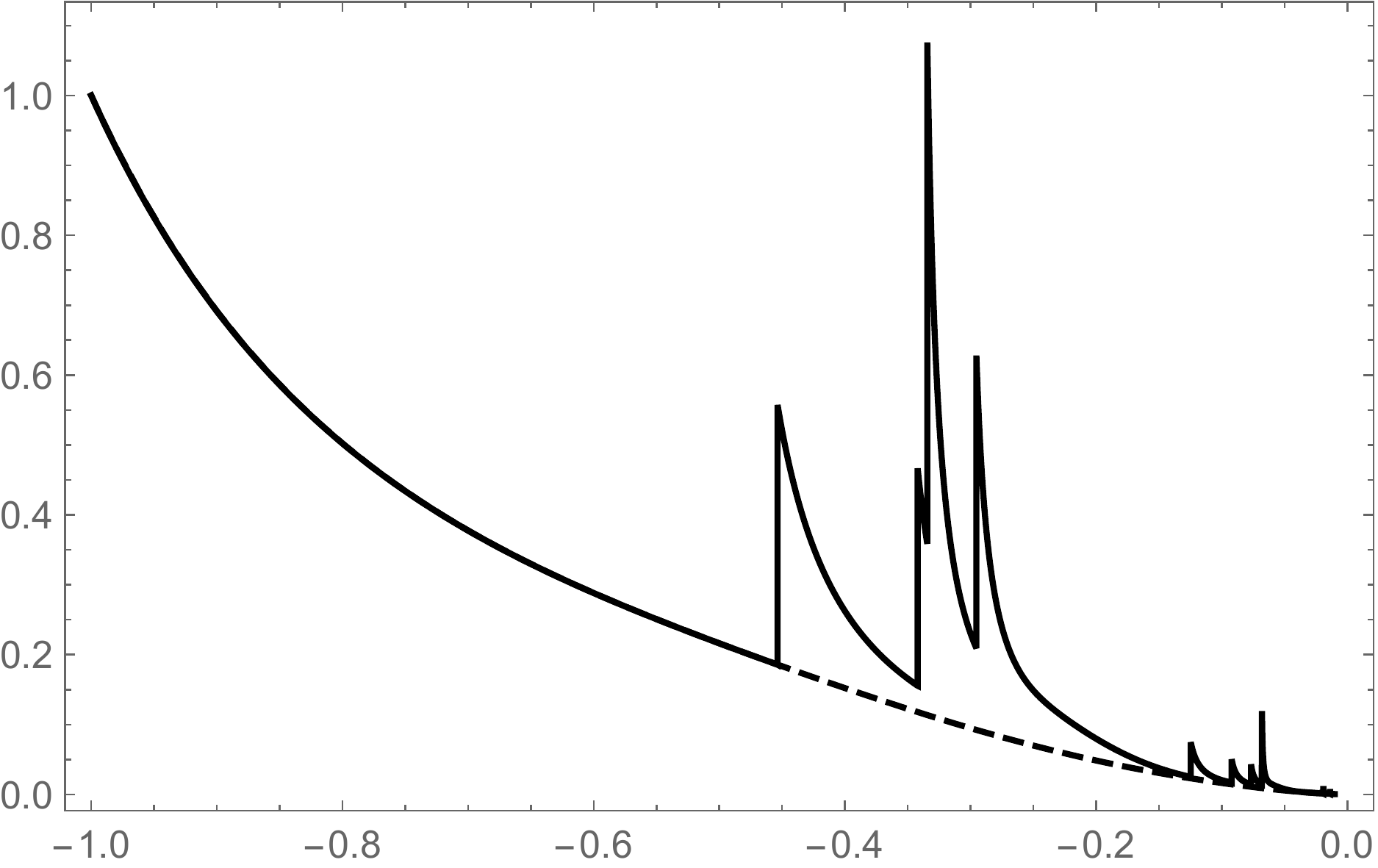}{p3}
\pgfdeclareimage[interpolate=true,width=10cm]{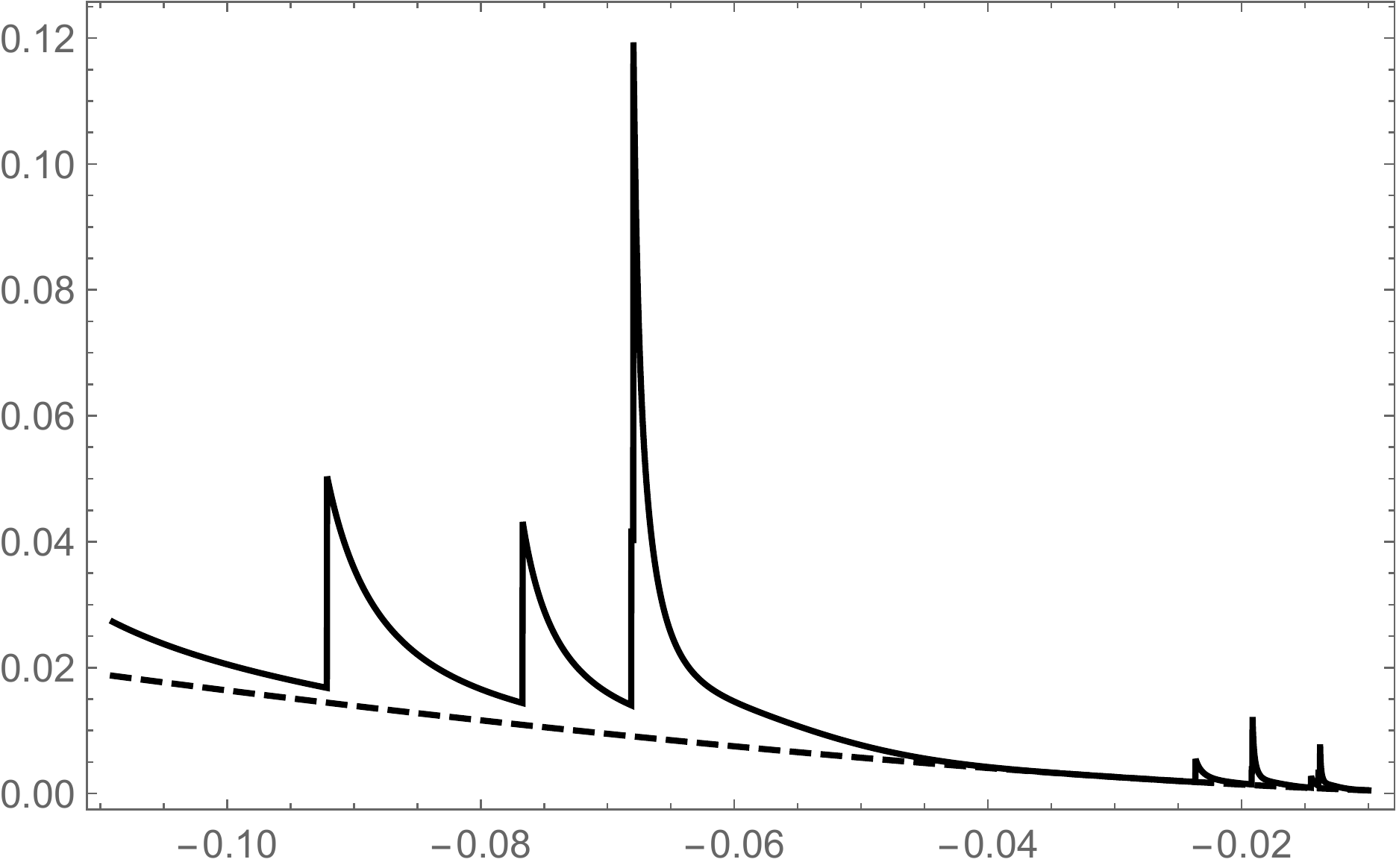}{p4}
\begin{figure}[!h]
\begin{center}
\begin{tikzpicture}[scale=0.73]
\pgftext[at=\pgfpoint{0cm}{0cm},left,base]{\pgfuseimage{p3}} 
\node at (5.18,-0.5) {\small$\tau$};
\node[rotate=90] at (-0.5,3.3) {\small$V_\vv$};
\begin{scope}[xshift=12cm]
\pgftext[at=\pgfpoint{0cm}{0cm},left,base]{\pgfuseimage{p4}} 
\node at (5.18,-0.5) {\small$\tau$};
\node[rotate=90] at (-0.5,3.3) {\small$V_\vv$};
\end{scope}
\end{tikzpicture}
\caption{\footnotesize  One simulation showing the behaviour of the variance $V_\vv$ including jumps (the dotted line shows the deterministic evolution) for $p=5$. The right figure is a close up of the small $|\tau|$ region. Notice that after a jump the variance rapidly relaxes to the underlying deterministic value.
It is clear that the jumps do not change the underlying behaviour of the deterministic evolution of $V_\vv$ as the mode exits the horizon.}
\label{fig4} 
\end{center}
\end{figure}
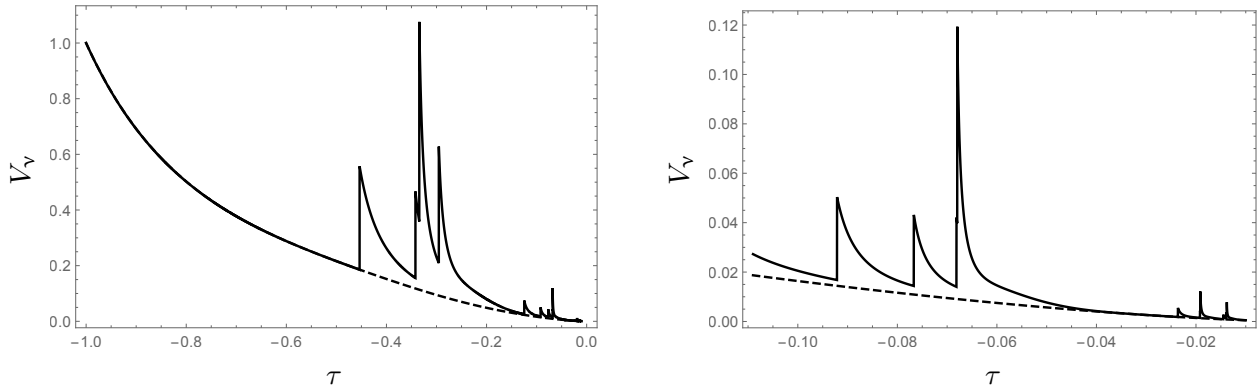

Up till now we have not considered the stochastic jumps and so now we turn to them. They occur with a rate, as $\tau\to\tau_\text{end}$, estimated via the Gaussian wave packet \eqref{gcs}, which scales as
\EQ{
\hat r=\sigma^2\hat V_\vv\thicksim\begin{cases}|\tau|^{(3-p)/2} & p>5\ ,\\[8pt] |\tau|^{-1} & p\leq5\ .\end{cases}
\label{rrf}
}
To start with, let is consider the behaviour of the variances when the jumps are included. A simulation of this is shown in figure \ref{fig4}. This is based on an approximation where the wave packet is assumed to be approximately Gaussian when the jumps occur. It is clear from this fairly crude analysis, that the jumps do not affect the overall localization of a mode, relative to the unconditioned state, as it exits the horizon. 
Hence, the degree of localization of a particular mode can be quantified, at the end of inflation, by the ratio
\EQ{
V_\vv\big|_\text{\tiny conditioned}\Big/V_\vv\big|_\text{\tiny unconditioned}\thicksim\begin{cases}e^{-\Delta N(p+1)/2} & p>5\ ,\\[8pt] e^{-\Delta N(p-2)} & p\leq5\ ,\end{cases}
}
where we have used $k|\tau_\text{end}|=e^{-\Delta N}$ at the end of inflation.
So for the modes that are relevant for the CMB, which exited the horizon $\Delta N\sim 50$ $e$-folds before the end of inflation, say, their conditioned state becomes extraordinarily narrow in field space and, hence, a classical description is entirely reasonable!

Now we consider the motion of the centre of the wave packet in phase space. From what we said above, this should a good coarse grained description when $p>5$ and for later times when the conditioned state has becomes effectively point-like in phase space. The special case $p=5$, realized in concrete models, is marginal in this regard and will require a more in-depth analysis than we present here.
When the wave packet jumps, the Gaussian form is not maintained and it is split into two wave packets. The non-linear deterministic dynamics \eqref{esh} then takes over and one of the offspring is amplified while the other fades away. The one that survives and the time it takes for the process depends on the detailed non-Gaussian form of the initial wave packet \cite{SH}. Following \cite{SH}, if one assumes that the relaxation occurs over a fast time scale, then one can coarse grain the process by describing the net effect of a jump to be a shift in position of the wave packet in phase space $(\bar\vv,\bar\uppi)$ by the form
\EQ{
\delta\bar\vv\approx\pm\sqrt{2\hat V_\vv}\ ,\qquad \delta\bar\uppi\approx\pm\sqrt{2\hat V_\uppi}\ ,
\label{nnb}
}
occurring with equal rate $\hat r/2$.
Since the rate of jumps $\hat r$ grows as $\tau\to\tau_\text{end}$ \eqref{rrf}, the effective process \eqref{nnb} looks more and more like a random walk with variances $\sigma_\vv=\sqrt{2\hat r\hat V_\vv}$ and $\sigma_\uppi=\sqrt{2\hat r\hat V_\uppi}$ in the $\vv$ and $\uppi$ direction. Therefore, including the deterministic evolution, the coarse grained dynamics of the position of the wave packet satisfies the Langevin equation
\EQ{
\frac{d\bar\vv}{d\tau}=\bar\uppi+\sigma_\vv\,\xi\ ,\qquad\frac{d\bar\uppi}{d\tau}=-\omega^2\bar\vv+\sigma_\uppi\,\xi\ ,
\label{lgn}
}
where $\xi(\tau)$ is a random Gaussian variable (white noise) with stochastic correlators
\EQ{
\mathscr E\big(\xi(\tau)\big)=0\ ,\qquad \mathscr E\big(\xi(\tau)\xi(\tau')\big)=\delta(\tau-\tau')\ .
}

We can pin down the variances $\sigma_\vv$ and $\sigma_\uppi$ by noticing that the probability density $P(\bar\vv,\bar\uppi)$ associated to the Langevin equation \eqref{lgn} satisfies the Fokker-Planck equation
\EQ{
\frac{\partial P}{\partial\tau}=-\bar\uppi\frac{\partial P}{\partial\bar\vv}+\omega^2\bar\vv\frac{\partial P}{\partial\bar\uppi}+\frac{\sigma_\uppi^2}{2}\frac{\partial^2P}{\partial\bar\uppi^2}\ ,
\label{fpe}
}
where we have not included the $\sigma_\vv$ term as it goes like $|\tau|^0$ and becomes subleading. Since, at the coarse grained level, for $p>5$, the conditioned state is effectively point-like in phase space and the unconditioned state is very decoherent (e.g.~has large entanglement entropy), the Wigner function of the unconditioned state 
acts as a probability density for the conditioned state in phase space. This means that we can identify $P$ with ${\cal W}$. Indeed, the master equation \eqref{mse} written in terms of the Wigner function of the unconditioned state is precisely of the form \eqref{fpe}. This identifies the coarse grained quantity $\sigma_\uppi=\sigma$ and also proves that the term involving $\sigma_\vv$ is, indeed, sub-leading, as anticipated. We can check the assumptions that have gone into the derivation of the Langevin equations by the following separate argument. The asymptotic form \eqref{dcc}, for $p\geq5$, shows that, as $\tau\to\tau_\text{end}$,  $\sigma_\uppi^2=2\hat r\hat V_\uppi=2\sigma^2\hat V_\vv\hat V_\uppi=\sigma^2$.

\pgfdeclareimage[interpolate=true,width=10cm]{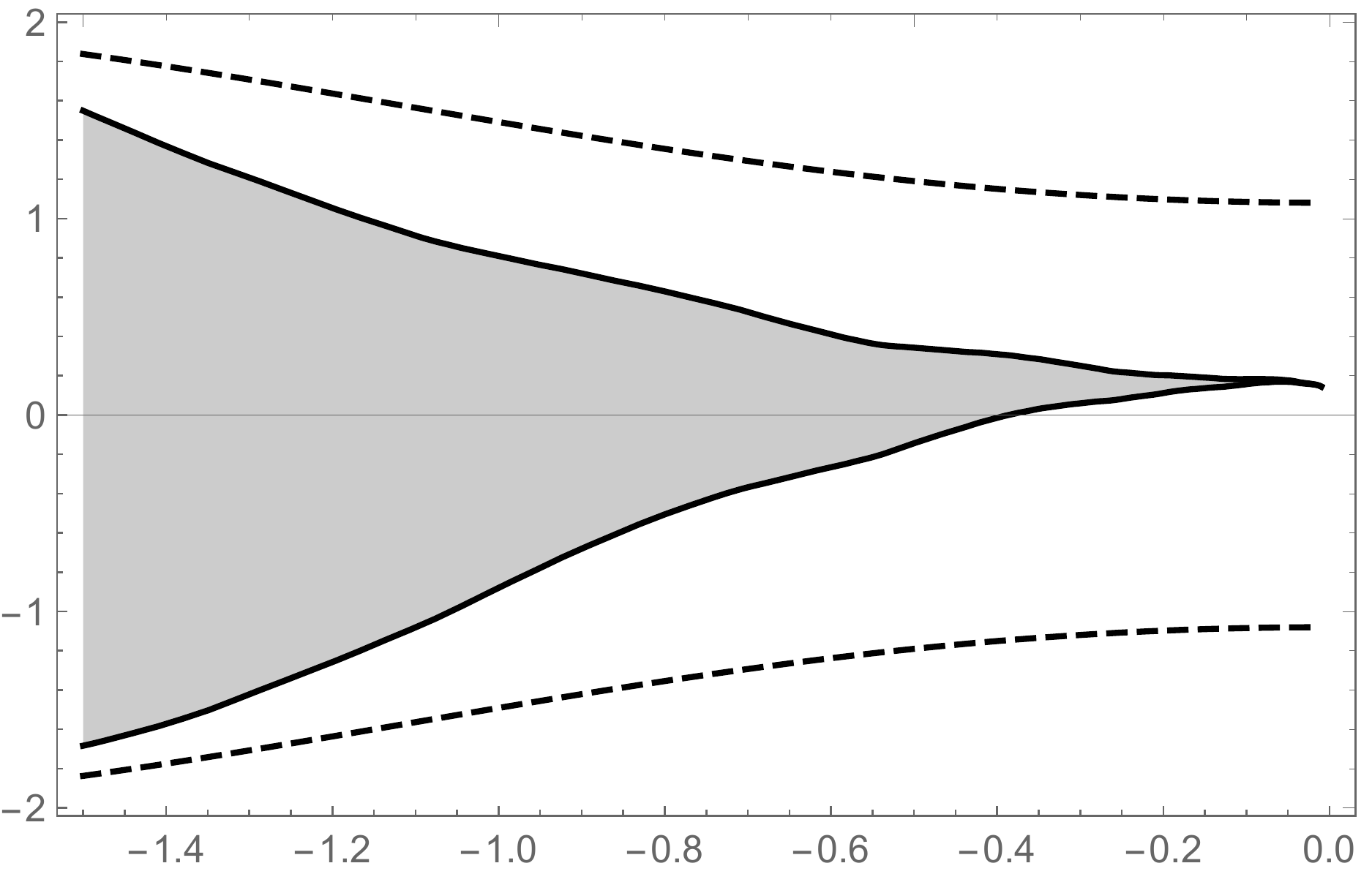}{p5}
\pgfdeclareimage[interpolate=true,width=10cm]{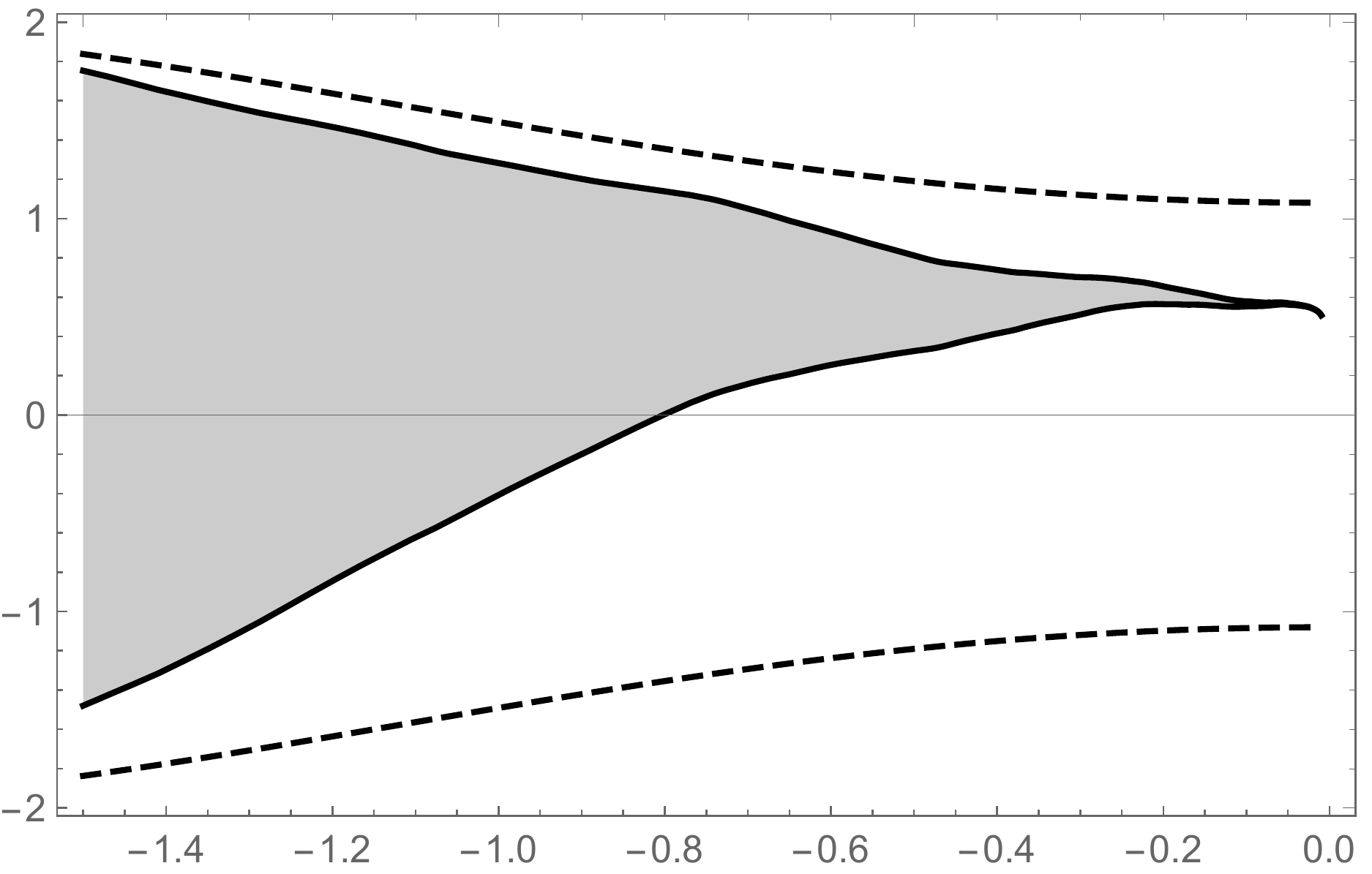}{p6}
\pgfdeclareimage[interpolate=true,width=10cm]{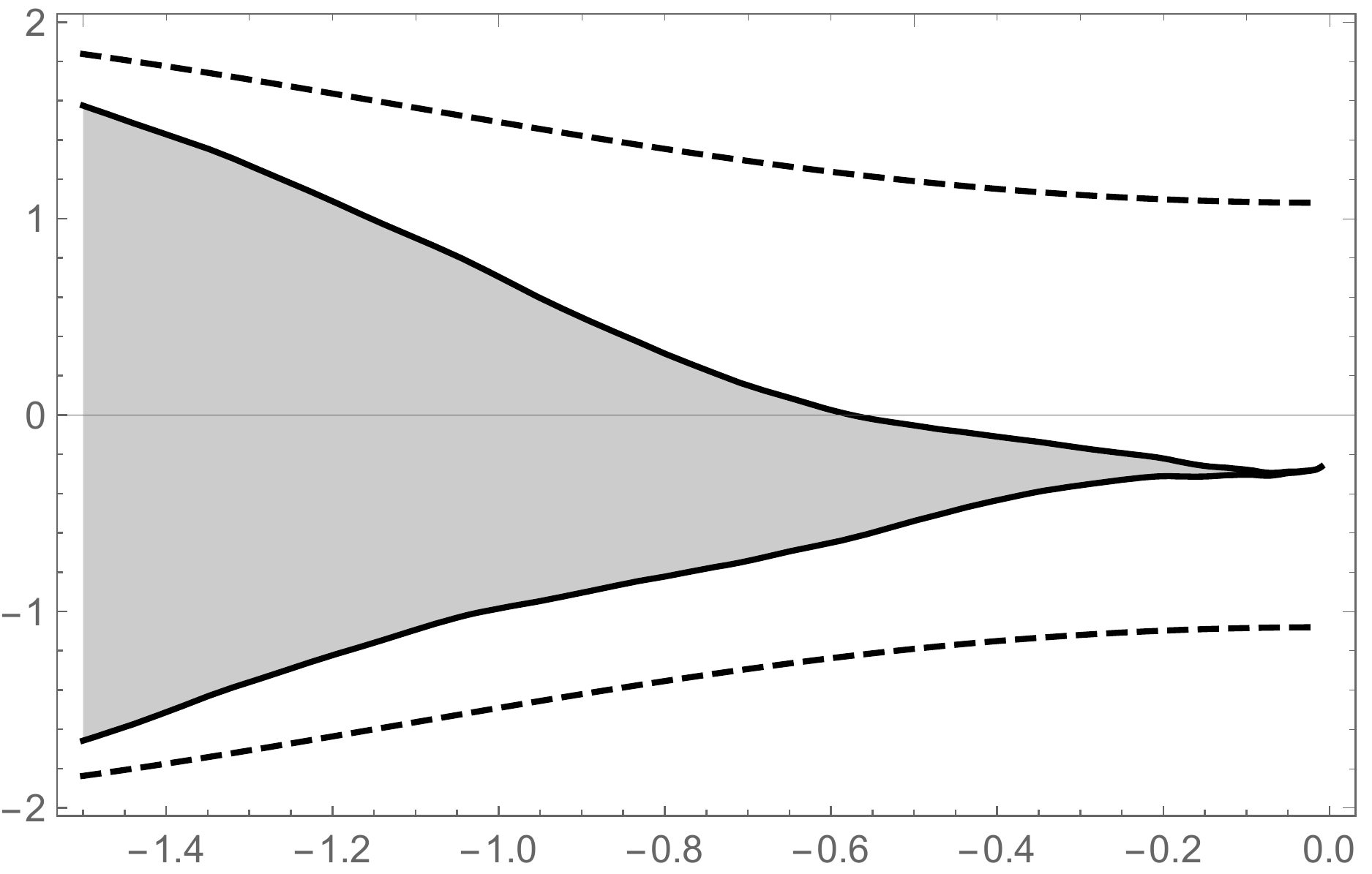}{p7}
\pgfdeclareimage[interpolate=true,width=10cm]{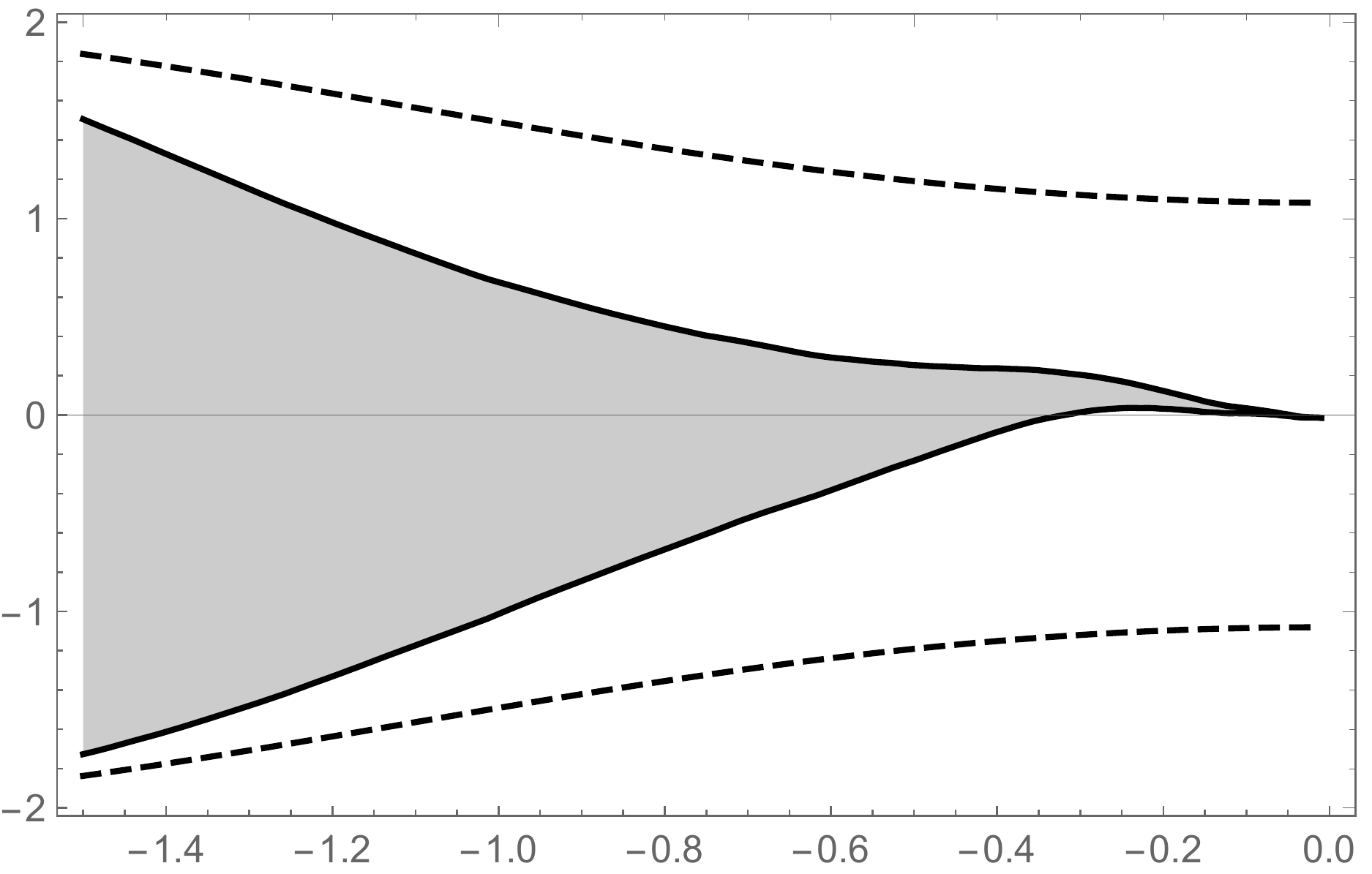}{p8}
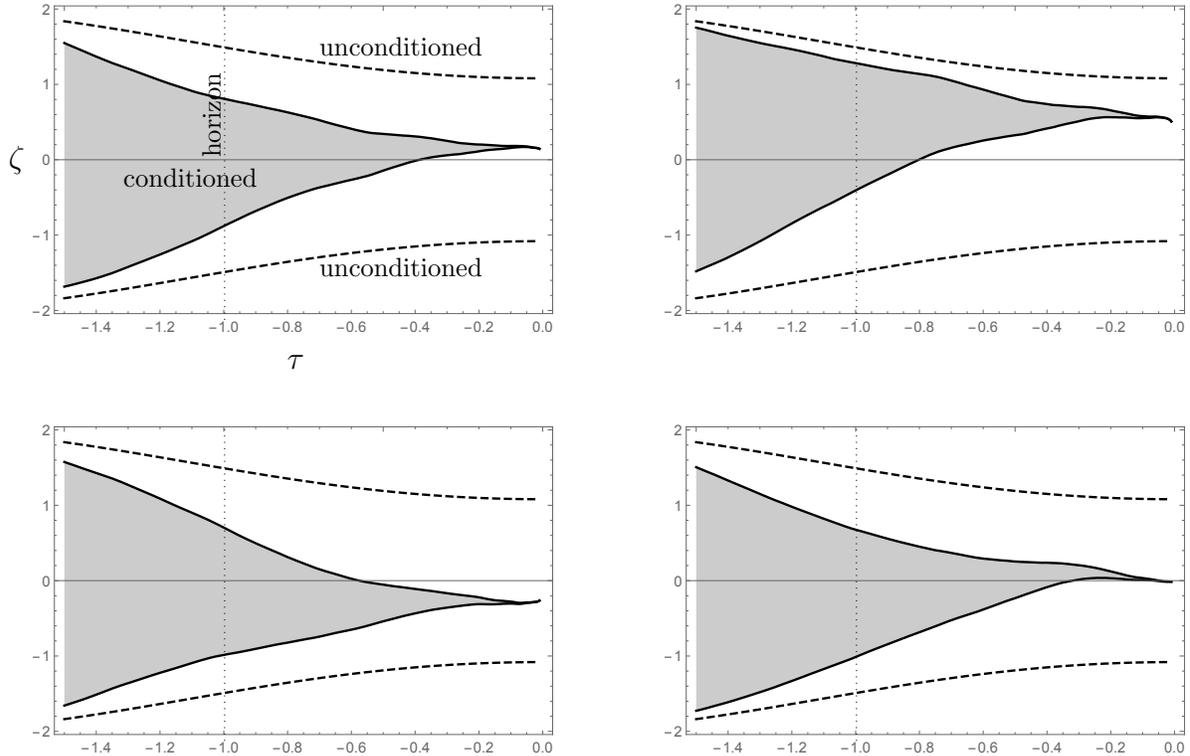
\begin{figure}[!h]
\begin{center}
\begin{tikzpicture}[scale=0.7]
\pgftext[at=\pgfpoint{0cm}{0cm},left,base]{\pgfuseimage{p5}} 
\draw[dotted] (3.65,0.3) -- (3.65,6.3);
\begin{scope}[xshift=12cm,yshift=0cm]
\pgftext[at=\pgfpoint{0cm}{0cm},left,base]{\pgfuseimage{p6}} 
\draw[dotted] (3.65,0.3) -- (3.65,6.3);
\end{scope}
\begin{scope}[xshift=0cm,yshift=-8cm]
\pgftext[at=\pgfpoint{0cm}{0cm},left,base]{\pgfuseimage{p7}} 
\draw[dotted] (3.65,0.3) -- (3.65,6.3);
\end{scope}
\begin{scope}[xshift=12cm,yshift=-8cm]
\pgftext[at=\pgfpoint{0cm}{0cm},left,base]{\pgfuseimage{p8}} 
\draw[dotted] (3.65,0.3) -- (3.65,6.3);
\end{scope}
\node at (7,1.3) {\footnotesize unconditioned}; 
\node at (7,5.5) {\footnotesize unconditioned}; 
\node at (3,3) {\footnotesize conditioned}; 
\node[rotate=90] at (3.4,4.2) {\footnotesize horizon}; 
\node at (5,-0.5) {$\tau$};
\node at (-0.3,3.35) {$\zeta $};
\end{tikzpicture}
\caption{\footnotesize  The evolution of the conditioned state (the dotted lines) and 4 realizations of the conditioned state (the shaded region) for the curvature perturbation $\zeta =\vv/(a\sqrt{2\varepsilon})$ in units of $H/\sqrt{2\varepsilon}$, with $p=6$. Note that the unconditioned state has a variance for $\zeta $ that freezes after exiting the horizon while the conditioned state clearly localizes.}
\label{fig5} 
\end{center}
\end{figure}

Four simulations of the conditioned state are shown in figure \ref{fig5} which plots the curvature perturbation $\zeta $ for the unconditioned and conditioned state. The picture is that the quantum-to-classical transition is a dynamical process that happens continuously. However, one can subjectively identify some $\tau$, say $\tau_c$, when the conditioned state becomes sufficiently localized relative to one's resolution scale, that the Langevin equation becomes a good description of the resulting coarse grained dynamics. The unconditioned state $\rho(\tau_c)$ then effectively provides stochastic initial conditions for the classical dynamics. 

It is interesting that the coarse grained stochastic dynamics of $(\bar\vv,\bar\uppi)$ is the same as that resulting from the quantum state diffusion unravelling of the master equation. This suggests that quantum state diffusion can act as a more tractable effective description of Born unravelling.

\section{Discussion}

Let us summarize the final picture we have established. The perturbations, like any subsystem of a bigger quantum system, define their own frame of reference. Their intrinsic state, the conditioned state, is always pure and evolves stochastically according to Born's rule. This evolution involves an effective non-Hermitian Hamiltonian that has the effect of localizing the state in field space. In every time interval there is also a chance that it jumps into a new branch of the unconditioned state $\ket{\psi}\to J\ket{\psi}$ which is orthogonal to the instantaneous conditioned state. A stochastic average of the conditioned state gives back the unconditioned state. Even as the unconditioned state spreads out as the mode exits the horizon, the conditioned state becomes more and more localized in field space, dependent on the exact model. Hence, the intrinsic state of the perturbations effectively becomes classical, specified by a point in phase space. An individual simulation of the conditioned state looks like a random walk. Modes with different wave vectors $\Bk$ then provide an ensemble of simulations whose statistics is governed by the unconditioned state.  One implication of the formalism is that as they exit the horizon thy modes do not completely freeze rather they are subject to random kicks from the environment described by a Langevin equation.

Finally, we believe that our formalism has implications for the stochastic inflation formalism \cite{Star1,SY}. In the latter, one defines a coarse grained field with a time dependent cut off that includes super Hubble modes with $k<\epsilon Ha$, for $\epsilon\ll1$ (i.e.~$k<\epsilon/|\tau|$). So new modes are continuously being included in the coarse grained field giving rise to a noise term in its equation of motion. In the present formalism, as modes are added to the coarse grained field they have been localized by a factor $\epsilon^{(p+1)/2}$, $p>5$ and $\epsilon^{p-2}$, $p\leq5$. So for $\epsilon\ll1$, it is consistent to treat the newly added modes as effectively classical.

\vspace{0.5cm}
\begin{center}{\scriptsize *************************}
\end{center}
\vspace{0.5cm}

I would like to thank Gianmassimo Tasinato, Jamie McDonald and Prem Kumar for comments and discussions. This work was partly supported by STFC grant ST/P00055X/1.

\end{document}